# Enhanced Second-Harmonic Generation in Thin-Film Lithium Niobate Circular Bragg Nanocavity

Zengya Li[1], Zhuoran Hu[1], Xiaona Ye[1], Zhengyang Mao[1], Juan Feng[1], Hao Li[1], Shijie Liu[1], Bo Wang[1], Yuanlin Zheng[1,2,4] and Xianfeng Chen[1,2,3,5]

[1]State Key Laboratory of Advanced Optical Communication Systems and Networks, School of Physics and Astronomy, Shanghai Jiao Tong University, Shanghai 200240, China

[2]Shanghai Research Center for Quantum Sciences, Shanghai 201315, China

[3]Collaborative Innovation Center of Light Manipulation and Applications, Shandong Normal University, Jinan 250358, China

[4]ylzheng@sjtu.edu.cn

[5]xfchen@sjtu.edu.cn

**Abstract:** Second-order nonlinearity gives rise to many distinctive physical phenomena, e.g., second-harmonic generation, which plays an important role in fundamental science and various applications. Lithium niobate, one of the most widely used nonlinear crystals, exhibits strong second-order nonlinear effects 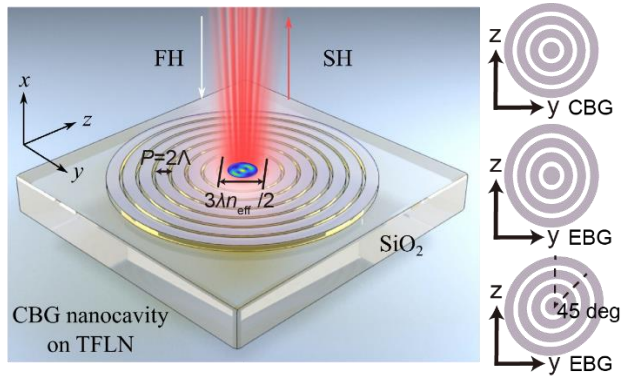 and electro-optic properties. However, its moderate refractive index and etching sidewall angle limit its capability in confining light into nanoscales, restricting its application in nanophotonics. Here, we exploit nanocavities formed by second-order circular Bragg gratings, which support resonant anapole modes to achieve highly enhanced SHG in thin film lithium niobate. The CBG nanocavity exhibits a record-high normalized conversion efficiency of $1.21 \times 10^{-2}$ cm$^2$/GW under the pump intensity of $1.9\,\mathrm{MW/cm^2}$. An SHG enhancement of 42,000 is realized compared to TFLN. Besides, we also show s- and p-polarization independent SHG in elliptical

Bragg nanocavities. This work could inspire studying nonlinear optics at the nanoscale on TFLN as well as other novel photonic platforms.

1. Introduction

Second-order nonlinear optical (NLO) responses are the core of diverse applications in nonlinear photonics, optical microscopy, quantum technology, and sensing.[1–3] Electro-optic effect, second-harmonic generation (SHG), spontaneous parametric down-conversion (SPDC), optical parametric oscillation and amplification (OPO/OPA), which are all classic examples of second-order NLO effects, have been widely used in electro-optic modulation,[4,5] frequency conversion,[6–8] quantum sources,[9] and single-molecule detection.[10] Second-order nonlinearity exists at the interface of media due to spatial inversion asymmetry.[11,12] The effect can be relatively strong for metals since surface plasmonic resonance provides strong light confinement. However, the high ohmic loss and the low overall nonlinear conversion efficiency have limited their practical use.[13,14] Dielectric materials without inversion symmetry, such as III-V semiconductors,[15–17] transition metal dichalcogenides (TMD),[18–21] and nonlinear crystals,[22–24] typically exhibit strong intrinsic second-order susceptibility. At the nanoscales, light-matter interaction length is much shorter than the coherent buildup wavelength ($L_{\text{coh}}$), thus phase-matching condition can be loosened. The nonlinear conversion efficiency highly depends on the local electromagnetic field strength, mode overlap, and nonlinear susceptibility.[25] Nanostructures such as nanoantenna, metasurfaces, and microcavities, which confine light into a small volume with an extended lifetime, have been demonstrated to significantly enhance light-matter interactions in the nanoscales.[26,27]

In recent years, thin film lithium niobate (TFLN) has emerged as a promising platform for densely integrated photonics. Lithium niobate (LN) is optically transparent ranging from ultraviolet to mid-infrared (0.4-5 μm) with a moderate refractive index ($n_o = 2.21$ and $n_e = 2.14$ at 1550 nm), strong second-order nonlinearity ($d_{33} = -27$ pm/V, $d_{31} = -4.3$ pm/V) and electro-optic effect ($\gamma_{51} = 32$ pm/V). The excellent properties of LN are widely used in electro-optic modulators, frequency converters, and nonlinear metasurfaces.[22,28] Devices on TFLN have shown unprecedented superior performance over bulk LN-based counterparts. Boosting SHG in nanostructure is critical for nanophotonics on the novel platform. To date, photonic crystal cavities,[4,29–31] plasmonic hybrid structures,[32–35] guide mode resonance structures,[36] and bound state in the continuum (BIC) structures[28,37] have been shown to enhance the nonlinear optical effects on TFLN. However, the low refractive index contrast between LN and the substrate, combined with non-perpendicular etching, leads to mode leakage, resulting in relatively low SHG conversion efficiency at the nanoscale.

Benefiting from their high light collection efficiency and vertical surface emission, circular Bragg grating (CBG) resonators have wide applications for lasers,[38] quantum emitters,[39,40] and nonlinear frequency converters.[41] Besides, the rotational symmetry of CBG is expected to realize polarization-independent performance.[42] The elliptical Bragg grating (EBG) can split the resonance modes and increase the linearly polarized photon collection efficiency.[43] However, most of the NLO resonators based on CBG and EBG utilize metallic structures with a low NLO conversion efficiency. Furthermore, CBG and EBG resonances in birefringence nonlinear crystals with complex polarization modes have not yet been explored.

Here, we achieve significantly enhanced SHG on CBG nanocavities based on x-cut TFLN. The experimental SHG conversion efficiency is measured to be $2.32 \times 10^{-5}$ with a 220-ps laser pump. The achieved normalized conversion efficiency is $1.21 \times 10^{-2}$ cm$^2$/GW under the pump intensity of $1.9$ MW/cm$^2$, with an enhancement factor of $4.2 \times 10^4$ compared to TFLN. Moreover, a 45-degree rotated elliptical Bragg grating (EBG) nanocavity with s-/p-polarization-independent SHG is experimentally demonstrated. The results show that the EBG nanocavity exhibits excellent polarization control performance of light, offering the potential to achieve polarization-independent devices.

2. **Results**

**Design and Fabrication**

The CBG nanocavities are designed and fabricated on a 300-nm-thick x-cut TFLN bonded on top of a 2-µm-thick silica buffering layer. There are two orthogonally polarized eigen resonance modes in the cavity due to the birefringence of LN: one polarization along the z-axis utilizing $d_{33}$, and the other along the y-axis utilizing $d_{31}$ of LN for SHG. The schematic of the CBG-enhanced SHG on TFLN is shown in Figure 1a. CBG satisfies the Bragg condition: $\lambda = 2n_{\text{eff}}\Lambda$, where $\lambda$, $n_{\text{eff}}$, and $\Lambda$ are the resonance wavelength, effective refractive index, and grating period, respectively. However, the non-perpendicular etching of LN increases the tolerances of the sample, resulting in poor performance of first-order CBG. Schematic diagram of first- and second-order CBG-enhanced anapole resonance are shown in Figure 1b. Anapole resonance enhancement is achieved by adjusting the size of the central nano-disk to $3\lambda n_{\text{eff}}/2$. CBG period, $P = 2\Lambda$, is designed based on second-order CBG condition. Here, we fabricated high-performance second-order CBGs (SEM image is shown in Figure 1c) which support anapole modes. The localized field of the fundamental harmonic (FH) is shown in Figures 1d-f, which is strongly localized in all directions and mostly confined in the center nano-disk.

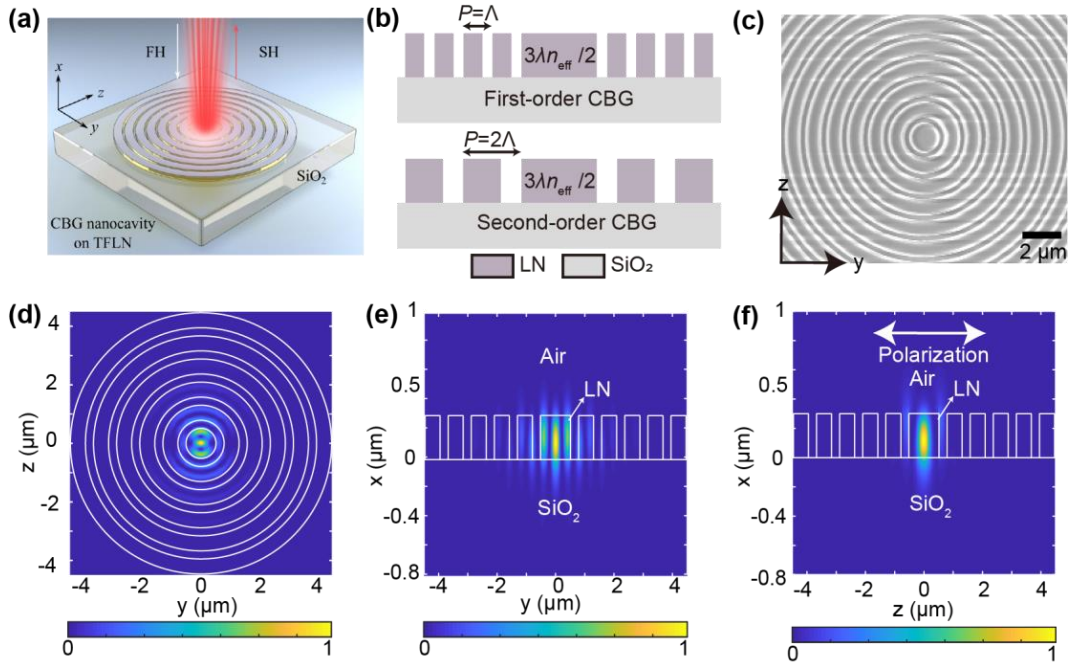

**Figure 1. Schematic of CBG resonance on TFLN.** (a) Schematic of CBG nanocavity on TFLN

for enhanced SHG. (**b**) Schematic diagram of first- and second-order CBG. (**c**) SEM image of CBG nanocavity. (**d-f**) YZ, XY, and XZ cross-sectional view of the resonant anapole mode intensity profile.

The reflection spectra (See Methods) of the CBG (periods ranging from 740 to 840 nm) nanocavity on TFLN are characterized using a supercontinuum laser in the NIR range. Details of the measurement setup are given in Methods. Figures 2a and 2b show the reflection spectra of CBG with different $P$'s under z-polarized and y-polarized illumination, respectively. The resonance is observed in the wavelength range of 1250-1450 nm for samples with different periods. The refractive index of $e$ wave (z-polarized illumination) is smaller than that of $o$ wave (y-polarized illumination) in LN, causing a redshift in the resonance. The full width at half maximum (FWHM) of anapole enhanced CBG is 5 nm, with an experimental Q factor of ~260 ($Q \approx \lambda/\Delta\lambda$). Figures 2c and 2d show the simulated reflection spectra corresponding to Figures 2a and 2b, respectively. The envelope in the experimental reflection curves is attributed to interference at the upper and lower interfaces of the TFLN layer, which isn't taken into consideration in the simulation.

Figures 2e and 2f show the reflection spectra of CBGs with different duty ratios ($D$) under z-polarized and y-polarized illumination, respectively. The effective refractive index of the grating increases as the duty increases, resulting in a resonance redshift. However, the FWHM increases and the $Q$ factor decreases due to mode leakage. Simulated reflection spectra are shown in Figures 2g and 2h, respectively.

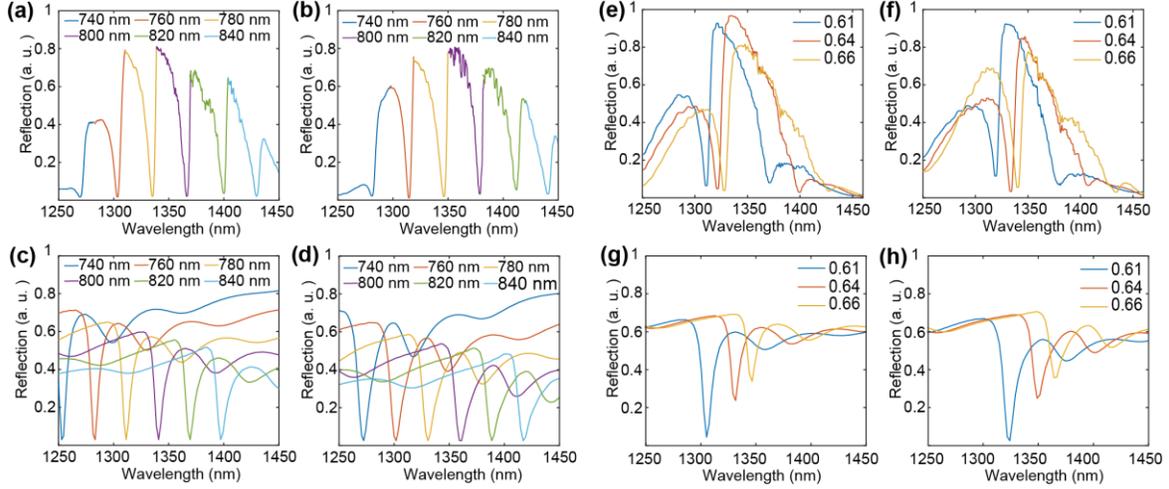

**Figure 2. Reflection characterization of the CBG nanocavities.** (**a**), (**b**) Reflection spectra of CBG with different $P$'s under z-polarized and y-polarized illumination. The simulated results are correspondingly (**c**) and (**d**). (**e**), (**f**) Reflection spectra of CBG with different $D$'s under z-polarized and y-polarized illumination. The simulated results are correspondingly (**g**) and (**h**).

## SHG characterization of CBG

Nonlinear optics characterization is performed on the fabricated samples using a home-built optical setup, shown in the Supplementary Materials. The SHG signals are always anisotropic, decided by the TFLN's asymmetry and orientation. The y- and z-component nonlinear polarization can be written as:

$$P_y = \varepsilon_0\left(-d_{22}E_x^2 + d_{22}E_y^2 + 2d_{31}E_yE_z\right),$$

$$P_z = \varepsilon_0\left(d_{31}E_x^2 + d_{31}E_y^2 + d_{33}E_z^2\right),$$

where $d_{22} = d_{21} = 2.1$ pm/V, $d_{31} = -4.3$ pm/V, and $d_{33} = -27$ pm/V, $E$ is the electric field, and $\varepsilon_0$ is the vacuum permittivity. For each resonance mode (either y-polarized or z-polarized), the SHG is anisotropic. A polar plot of the SHG signal on the polarization of the resonant FH wavelengths of 1302 nm and 1314 nm is shown in Figures 3a and 3b, respectively. Although the CBG structure is isotropic, its resonance and the SHG signal still exhibit anisotropy due to the birefringence effect of LN. SHG signal spectra for CBG nanocavities with different $P$'s and $D$'s under z-polarized illumination are shown in Figures 3c and 3d. With the increasing of $P$ and $D$, the resonant wavelength redshifts, so as the SH spectra. The FWHM of the SH spectra remains similar, indicating that the Q factors of each CBG nanocavity are comparable.

The pump light at the central wavelength of 1302 nm with an FWHM of 5 nm is used to measure the nonlinear conversion efficiency of the CBG nanocavity. The intensity of the SHG shows a good quadratic relationship (Figure 3e) with that of the FH pump incident with a pulse width of 220 ps (Figure S3). In our experiment, SH conversion efficiency $\eta = P_{SH}/P_{FH}$ of $2.32 \times 10^{-5}$ is achieved with the averaged pump power of 0.86 mW, corresponding to a peak intensity ($I_{peak}$) of 1.9 MW/cm$^2$. The normalized efficiency ($\eta_{norm} = \eta/I_{peak}$) of CBG nanocavity is calculated to be $1.21 \times 10^{-2}$ cm$^2$/GW. The results have exceeded those in the guided-mode-resonance and membrane metasurface[36,44] by three orders of magnitude.

The SHG enhancement factor varying with periods and duties (Figure 3f) is calculated by integrating the intensity of the CBG resonant spot and that of flat TFLN under the same focused pump (See Methods). In the optimized CBG nanocavity ($P = 760$ nm, $D = 0.66$), the maximum SHG enhancement factor reaches $4.2 \times 10^4$ under z-polarized FH illumination, surpassing previous achievements of TFLN nanostructure.[36] When $P$ is smaller, the resonance wavelength experiences a blueshift. The film interference at this wavelength leads to strong transmission and weak reflection (Figure 2a), significantly reducing the SHG enhancement factor. On the contrary, when $P$ is larger, the resonance wavelength undergoes a redshift, which is accompanied by mode leakage in the y-direction of the cavity, resulting in a slight decrease of the SHG enhancement factor. The anapole resonant-SHG spots of the sample (exposure time of 2 ms) under z- and y-polarized FH at 1302 nm and 1314 nm are shown in Figures 3g and 3h, respectively. Compared with the SHG spot (exposure time of 2000 ms) on flat TFLN under z-polarized FH (Figure 3i), a giant second harmonic enhancement is visible in CBG.

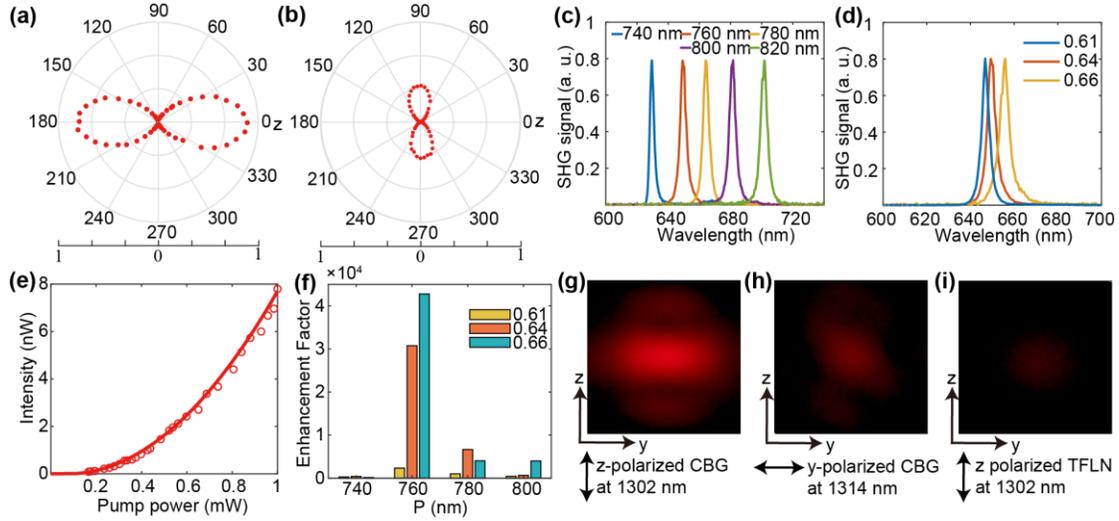

**Figure 3. SHG enhancement of CBG nanocavities on TFLN.** (**a**) 1302 nm utilizing $d_{33}$ and (**b**) 1314 nm utilizing $d_{31}$ of LN. (**c** and **d**) SHG spectra of CBG with different $P$'s and $D$'s under z-polarized illumination. (**e**) Quadratic relationship between SH and FH power with $P = 760$ nm and $D = 0.66$. (**f**) SHG enhancement factor varying with $P$ and $D$. (**g** and **h**) Z-and y-polarized resonant SHG spot of CBG. (**i**) Z-polarized SHG spot of flat TFLN.

The orientation of elliptical Bragg nanocavities in nonlinear crystals is attractive to modulating nonlinear signals. Here, resonant FH polarization independence has been achieved in CBG nanocavities, however, the nonlinear effect remains polarization-dependent. To tune the polarization of the SHG signal, the CBG cavity is elongated along the z-axis by 8 nm for every period to form an EBG one while keeping unchanged along y-axis (inset of Figure 4a), causing the resonance modes with z- and y-polarization to coalesce. Figure 4a illustrates the SHG power as a function of the pump polarization angle along z-axis at the wavelength of 1306 nm. There is no resonance wavelength shift for different polarization of FH, only a change in the intensity of the SHG (Figure 4b). This means that the birefringence of TFLN is well balanced by EBG nanocavity. SHG spots corresponding to the three typical polarized FH (0, 45, 90 degrees) in Figure 4b are shown in Figures 4c-e. The SHG intensity varies by three times during the whole polarization rotation. When the FH polarization rotates, the orientation of the SHG spot rotates accordingly. The spot orientation in Figure 4e does not follow the z-axis, indicating that SHG attributes to the combined effect of $d_{33}$, $d_{31}$ and $d_{22}$ nonlinearity.

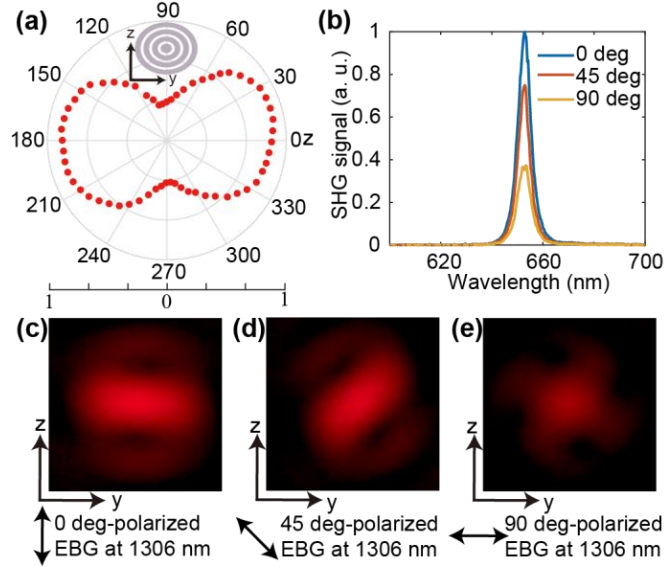

**Figure 4. SHG enhancement results of EBG.** (**a**) Dependence of SHG power under the resonant FH pump. (**b**) SH spectra for FH polarization at 0, 45, 90 degrees. (**c**) Z-polarized resonant SHG spot. (**d**) SHG spot of under 45 degree-polarized resonant FF. (**e**) Y-polarized resonant SHG spot.

Moreover, we demonstrate the s- and p-polarization (resonant modes at 1296 and 1304 nm) independent effect generated by a 45-degree rotated EBGs (inset of Figure 5a). The relationship between these two modes as a function of polarization angle is shown in Figures 5a and 5b. The two orthogonal polarizations are 66 and 156 degrees relative to z-axis. Unlike CBG, the SHG intensity of the two orthogonal s- and p-polarization modes only have a 20% difference. Although the rotation reduces the dependence on $d_{33}$, and decreases SHG signal intensity, it cancels out the collection by elliptical effect,[43] resulting in the collected intensity of the SHG signal not decreasing compared to the CBG nanocavity. When the resonant wavelength (1301 nm) is between the two orthogonal polarization modes, s- and p-polarization independence is shown in Figure 5c. The coherent superposition of the two modes results in the partial polarization independence for the s-polarized and p-polarized FH excitation. In the s- and p-polarization directions, the intensities are almost equal, and the maximum SHG intensity is only twice as much as the minimum. The SHG spectra excited by a supercontinuum light source are shown in Figure 5d. The spots at the corresponding polarization angle are shown in Figures 5e-g. Within the overlap range of the two modes, conversion of the two modes can be observed.

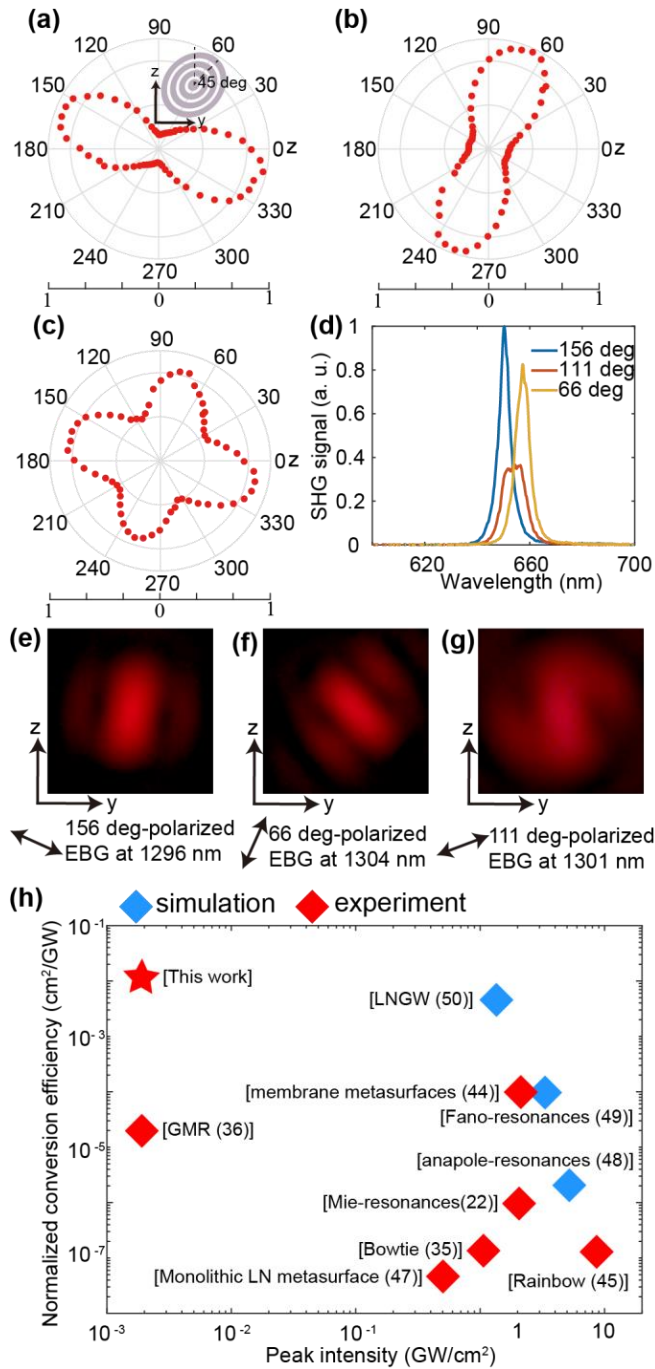

**Figure 5. SHG enhancement of EBG with a rotation angle of 45 degrees.** (**a-c**) Dependence of SHG enhancement on the resonant wavelength of 1296, 1304, and 1301 nm FH pump. (**d**) The spectrum of SHG at 66, 111, 156 degrees of polarization. (**e-g**) SHG spot at 66, 111, 156 degrees of polarization angle. (**h**) Comparison of SHG conversion efficiency of different structures.

## 3. Discussion

Nanocavities, characterized by its high Q-factor, strong light confinement, and efficient light collection, are widely utilized in quantum light source devices. We leverage CBG nanocavities on TFLN to enhance SHG and have surpassed the current conversion efficiency in the nano-scale. About $4.2 \times 10^4$ times enhancement factor is realized under resonant 220 ps laser excitation. The achieved nonlinear conversion efficiency reaches $2.32 \times 10^{-5}$ and the normalized efficiency of CBG nanocavity is calculated as $1.21 \times 10^{-2}$ cm$^2$/GW. A detailed comparison of CBG nanocavity with typical reports is shown in Figure 5h.

We have experimentally exceeded the normalized conversion efficiency of LN nanostructure to $10^{-2}$ cm$^2$/GW. Although the conversion efficiency of SHG can be enhanced by combining metal and LN,[32–35] this conversion efficiency is still limited. The ohmic loss, zero bulk second-order nonlinearity of metals, and the low damage threshold of plasmonic structures result in a low nonlinear conversion efficiency.[45] The moderate refractive index of LN, results in low Q factors of Mie resonance less than 100,[22,46,47] which limits its SHG efficiency. Fano and anapole resonance structures can theoretically increase normalized conversion efficiency to $10^{-5}$ cm$^2$/GW.[48,49] Representative development is that the GMR and membrane metasurface structures increase the normalized conversion efficiency to $10^{-5}$ cm$^2$/GW.[36,44] Although the lithium niobate grating waveguide (LNGW)[50] structure can theoretically achieve a conversion efficiency of $10^{-3}$ cm$^2$/GW, it is difficult to carry out due to the stringent fabrication conditions. We achieve the highest normalized conversion efficiency on TFLN under the lowest pump intensity utilizing CBG nanocavities. Besides, CBG is extended to EBG to control the polarization of SHG in nonlinear crystals and s- and p-polarization independence can be achieved without reducing the nonlinear conversion efficiency (order of $10^{-2}$ cm$^2$/GW). An insightful work is to add a gold reflective layer under TFLN to achieve electro-optically tunable nonlinear effects with a normalized conversion efficiency to the order of $10^{-1}$ cm$^2$/GW or even higher. Moreover, our method can be easily applied to the III-V semiconductor and 2D material platforms, and a new scheme is proposed for studying nonlinear optical effects at the nanoscale.

## 4. Conclusion

In summary, high-performance anapole resonance enhanced second-order CBG nanocavities are designed on TFLN. About $4.2 \times 10^4$ times SHG enhancement factor is realized under resonant 220-ps laser excitation. The nonlinear conversion efficiency reaches $2.32 \times 10^{-5}$ and the normalized efficiency of CBG nanocavity is estimated as $1.21 \times 10^{-2}$ cm$^2$/GW. Furthermore, we have also achieved SHG s- and p- polarization independence in TFLN based on EBG nanocavity. The scheme can also be extended to other nonlinear optical platforms, such as transition metal dichalcogenides and III-V semiconductors. This work provides a new approach for studying nonlinear optics at the nanoscale void of phase matching.

## METHODS

**Experimental setup**

The light source is a picosecond supercontinuum laser, with a pulse duration of 220 ps (YSL photonics SC-pro). The supercontinuum is firstly filtered by a long-pass filter (>1100 nm). The input light is then tightly focused on the sample by a NIR microscope objective (50×, NA=0.67). Reflection light is collected by the same objective and reflected by a beam splitter (50:50, 600-1700 nm). A flip mirror is added after the beam splitter to guide the beam into two arms. The transmitted light is coupled into a multimode fiber and recorded by a spectrometer or imaged using a camera. During the SHG characterization, the supercontinuum is filtered by a tunable light filter (YSL photonics AOTF-PRO2, FHWM 5 nm).

**Calculation of SH conversion efficiency**

The maximum SHG power reflected from the sample (FH at 1302 nm resonance) is firstly measured by a photodiode power meter (Thorlabs, S130C, 400-1100 nm) to be 20 nW. Two shortpass filters (OD>7) are used to filter the FH wave. The SHG intensity is then calibrated using the CCD (VIHENT, VTSE3-600). The power is mapped with the sum count of $N_{\max}$. The exposure parameter of 0.4 ms and the lowest gain of 100 are set. The CCD exposure time $t_{ex}$ and gain $G$ are linearly proportional to its count. Thus, the SHG signal with a sum count of $N$ is calibrated to be $P_{SH} = \frac{N}{t_{ex} \times G} / \frac{N_{max}}{0.4 \text{ ms} \times 100} \times 20 \text{ nW}$.

SHG conversion efficiency is defined as $\eta = P_{SH}/P_{FH}$ where $P_{SH}$ and $P_{FH}$ are the average powers of the SHG signal and FH power, respectively. The peak intensity $I_{peak}$ is calculated as $I_{peak} = P/f_{rep} \cdot \tau$, where $f_{rep}$ is the pulse repetition rate and $\tau = 220$ ps is the pulse duration. The normalized SHG conversion efficiency $\eta_{norm}$ is calculated as $\eta_{norm} = \eta/I_{peak}$.

**References:**

(1) Boyd, R. W.; Gaeta, A. L.; Giese, E. Nonlinear Optics. In *Springer Handbook of Atomic, Molecular, and Optical Physics*; Springer, 2008; pp 1097–1110.

(2) Xia, H.; Wu, S.; Liang, B.; Ding, J.; Jiang, Z.; Dong, G.; Shi, Y.; Shen, D.; Cheng, J.; Liu, W.-T. Nonlinear Optical Signatures of Topological Dirac Fermion. *Sci. Adv.* **2024**, *10* (25), eadp0575.

(3) Sultanov, V.; Kavčič, A.; Kokkinakis, E.; Sebastián, N.; Chekhova, M. V.; Humar, M. Tunable Entangled Photon-Pair Generation in a Liquid Crystal. *Nature* **2024**. https://doi.org/10.1038/s41586-024-07543-5.

(4) Vogler-Neuling, V. V.; Karvounis, A.; Morandi, A.; Weigand, H.; Dénervaud, E.; Grange, R. Photonic Assemblies of Randomly Oriented Nanocrystals for Engineered Nonlinear and Electro-Optic Effects. *ACS Photonics* **2022**, *9* (7), 2193–2203.

(5) Wang, C.; Zhang, M.; Chen, X.; Bertrand, M.; Shams-Ansari, A.; Chandrasekhar, S.; Winzer, P.; Lončar, M. Integrated Lithium Niobate Electro-Optic Modulators Operating at CMOS-Compatible Voltages. *Nature* **2018**, *562* (7725), 101–104.

(6) Li, Z.; Corbett, B.; Gocalinska, A.; Pelucchi, E.; Chen, W.; Ryan, K. M.; Khan, P.; Silien, C.; Xu, H.; Liu, N. Direct Visualization of Phase-Matched Efficient Second Harmonic and Broadband Sum Frequency Generation in Hybrid Plasmonic Nanostructures. *Light Sci. Appl.* **2020**, *9* (1), 180.

(7) Wang, J.-Q.; Yang, Y.-H.; Li, M.; Hu, X.-X.; Surya, J. B.; Xu, X.-B.; Dong, C.-H.; Guo, G.-C.; Tang, H. X.; Zou, C.-L. Efficient Frequency Conversion in a Degenerate χ (2) Microresonator. *Phys. Rev. Lett.* **2021**, *126* (13), 133601.

(8) Ling, J.; Staffa, J.; Wang, H.; Shen, B.; Chang, L.; Javid, U. A.; Wu, L.; Yuan, Z.; Lopez-Rios, R.; Li, M. Self-Injection Locked Frequency Conversion Laser. *Laser Photonics Rev.* **2023**, *17* (5), 2200663.

(9) Dutt, A.; Mohanty, A.; Gaeta, A. L.; Lipson, M. Nonlinear and Quantum Photonics Using Integrated Optical Materials. *Nat. Rev. Mater.* **2024**, *9* (5), 321–346. https://doi.org/10.1038/s41578-024-00668-z.

(10) Wang, H.; Lee, D.; Cao, Y.; Bi, X.; Du, J.; Miao, K.; Wei, L. Bond-Selective Fluorescence Imaging with

Single-Molecule Sensitivity. *Nat. Photonics* **2023**, *17* (10), 846–855.

(11) Wang, J.; Zhu, B.-F.; Liu, R.-B. Second-Order Nonlinear Optical Effects of Spin Currents. *Phys. Rev. Lett.* **2010**, *104* (25), 256601.

(12) Fu, X.; Zeng, Z.; Jiao, S.; Wang, X.; Wang, J.; Jiang, Y.; Zheng, W.; Zhang, D.; Tian, Z.; Li, Q. Highly Anisotropic Second-Order Nonlinear Optical Effects in the Chiral Lead-Free Perovskite Spiral Microplates. *Nano Lett.* **2023**, *23* (2), 606–613.

(13) Dong, Z.; Asbahi, M.; Lin, J.; Zhu, D.; Wang, Y. M.; Hippalgaonkar, K.; Chu, H.-S.; Goh, W. P.; Wang, F.; Huang, Z. Second-Harmonic Generation from Sub-5 Nm Gaps by Directed Self-Assembly of Nanoparticles onto Template-Stripped Gold Substrates. *Nano Lett.* **2015**, *15* (9), 5976–5981.

(14) Zhang, C.-C.; Zhang, J.-Y.; Feng, J.-R.; Liu, S.-T.; Ding, S.-J.; Ma, L.; Wang, Q.-Q. Plasmon-Enhanced Second Harmonic Generation of Metal Nanostructures. *Nanoscale* **2024**, *16* (12), 5960–5975.

(15) Liu, S.; Sinclair, M. B.; Saravi, S.; Keeler, G. A.; Yang, Y.; Reno, J.; Peake, G. M.; Setzpfandt, F.; Staude, I.; Pertsch, T. Resonantly Enhanced Second-Harmonic Generation Using III–V Semiconductor All-Dielectric Metasurfaces. *Nano Lett.* **2016**, *16* (9), 5426–5432.

(16) Timofeeva, M.; Lang, L.; Timpu, F.; Renaut, C.; Bouravleuv, A.; Shtrom, I.; Cirlin, G.; Grange, R. Anapoles in Free-Standing III–V Nanodisks Enhancing Second-Harmonic Generation. *Nano Lett.* **2018**, *18* (6), 3695–3702.

(17) Lu, X.; Moille, G.; Rao, A.; Westly, D. A.; Srinivasan, K. Efficient Photoinduced Second-Harmonic Generation in Silicon Nitride Photonics. *Nat. Photonics* **2021**, *15* (2), 131–136.

(18) Chowdhury, T.; Sadler, E. C.; Kempa, T. J. Progress and Prospects in Transition-Metal Dichalcogenide Research beyond 2D. *Chem. Rev.* **2020**, *120* (22), 12563–12591.

(19) Nauman, M.; Yan, J.; de Ceglia, D.; Rahmani, M.; Zangeneh Kamali, K.; De Angelis, C.; Miroshnichenko, A. E.; Lu, Y.; Neshev, D. N. Tunable Unidirectional Nonlinear Emission from Transition-Metal-Dichalcogenide Metasurfaces. *Nat. Commun.* **2021**, *12* (1), 5597.

(20) Shree, S.; Lagarde, D.; Lombez, L.; Robert, C.; Balocchi, A.; Watanabe, K.; Taniguchi, T.; Marie, X.; Gerber, I. C.; Glazov, M. M. Interlayer Exciton Mediated Second Harmonic Generation in Bilayer MoS2. *Nat. Commun.* **2021**, *12* (1), 6894.

(21) Yuan, Y.; Liu, P.; Wu, H.; Chen, H.; Zheng, W.; Peng, G.; Zhu, Z.; Zhu, M.; Dai, J.; Qin, S. Probing the Twist-Controlled Interlayer Coupling in Artificially Stacked Transition Metal Dichalcogenide Bilayers by Second-Harmonic Generation. *ACS Nano* **2023**, *17* (18), 17897–17907.

(22) Ma, J.; Xie, F.; Chen, W.; Chen, J.; Wu, W.; Liu, W.; Chen, Y.; Cai, W.; Ren, M.; Xu, J. Nonlinear Lithium Niobate Metasurfaces for Second Harmonic Generation. *Laser Photonics Rev.* **2021**, *15* (5), 2000521.

(23) Kang, L.; Lin, Z. Deep-Ultraviolet Nonlinear Optical Crystals: Concept Development and Materials Discovery. *Light Sci. Appl.* **2022**, *11* (1), 201.

(24) Zhang, J.; Ma, J.; Parry, M.; Cai, M.; Camacho-Morales, R.; Xu, L.; Neshev, D. N.; Sukhorukov, A. A. Spatially Entangled Photon Pairs from Lithium Niobate Nonlocal Metasurfaces. *Sci. Adv.* **2022**, *8* (30), eabq4240.

(25) Zhang, T.; Guo, Q.; Shi, Z.; Zhang, S.; Xu, H. Coherent Second Harmonic Generation Enhanced by Coherent Plasmon–Exciton Coupling in Plasmonic Nanocavities. *ACS Photonics* **2023**, *10* (5), 1529–1537.

(26) Chen, S.; Kang, E. S. H.; Shiran Chaharsoughi, M.; Stanishev, V.; Kühne, P.; Sun, H.; Wang, C.; Fahlman, M.; Fabiano, S.; Darakchieva, V.; Jonsson, M. P. Conductive Polymer Nanoantennas for Dynamic Organic Plasmonics. *Nat. Nanotechnol.* **2020**, *15* (1), 35–40.

(27) Churaev, M.; Wang, R. N.; Riedhauser, A.; Snigirev, V.; Blésin, T.; Möhl, C.; Anderson, M. H.; Siddharth, A.; Popoff, Y.; Drechsler, U.; Caimi, D.; Hönl, S.; Riemensberger, J.; Liu, J.; Seidler, P.; Kippenberg, T. J. A Heterogeneously Integrated Lithium Niobate-on-Silicon Nitride Photonic Platform. *Nat. Commun.* **2023**, *14* (1), 3499.

(28) Liu, S.; Hong, W.; Sui, X.; Hu, X. High-Efficiency Second-Harmonic Generation Using Quasi-Bound State in LiNbO3 Metasurface. In *Photonics*; MDPI, 2023; Vol. 10, p 661.

(29) Li, M.; Ling, J.; He, Y.; Javid, U. A.; Xue, S.; Lin, Q. Lithium Niobate Photonic-Crystal Electro-Optic Modulator. *Nat. Commun.* **2020**, *11* (1), 4123. https://doi.org/10.1038/s41467-020-17950-7.

(30) Jiang, H.; Liang, H.; Luo, R.; Chen, X.; Chen, Y.; Lin, Q. Nonlinear Frequency Conversion in One Dimensional Lithium Niobate Photonic Crystal Nanocavities. *Appl. Phys. Lett.* **2018**, *113* (2), 021104.

(31) Wei, D.; Wang, C.; Wang, H.; Hu, X.; Wei, D.; Fang, X.; Zhang, Y.; Wu, D.; Hu, Y.; Li, J. Experimental

Demonstration of a Three-Dimensional Lithium Niobate Nonlinear Photonic Crystal. *Nat. Photonics* **2018**, *12* (10), 596–600.

(32) Lehr, D.; Reinhold, J.; Thiele, I.; Hartung, H.; Dietrich, K.; Menzel, C.; Pertsch, T.; Kley, E.-B.; Tünnermann, A. Enhancing Second Harmonic Generation in Gold Nanoring Resonators Filled with Lithium Niobate. *Nano Lett.* **2015**, *15* (2), 1025–1030.

(33) Singh, G.; Kumar, M.; Singh, M.; Vaish, R. Surface Plasmon Resonance Triggered Promising Visible Light Photocatalysis of LiNbO3 Ceramic Supported Ag Nanoparticles. *J. Am. Ceram. Soc.* **2021**, *104* (3), 1237–1246.

(34) Ali, R. F.; Busche, J. A.; Kamal, S.; Masiello, D. J.; Gates, B. D. Near-Field Enhancement of Optical Second Harmonic Generation in Hybrid Gold–Lithium Niobate Nanostructures. *Light Sci. Appl.* **2023**, *12* (1), 99.

(35) Li, Z.; Ye, X.; Hu, Z.; Li, H.; Liu, S.; Zheng, Y.; Chen, X. Plasmonic Hotspot Arrays Boost Second Harmonic Generation in Thin-Film Lithium Niobate. *Opt. Express* **2024**, *32* (8), 13140–13155.

(36) Yuan, S.; Wu, Y.; Dang, Z.; Zeng, C.; Qi, X.; Guo, G.; Ren, X.; Xia, J. Strongly Enhanced Second Harmonic Generation in a Thin Film Lithium Niobate Heterostructure Cavity. *Phys. Rev. Lett.* **2021**, *127* (15), 153901.

(37) Yu, Z.; Tong, Y.; Tsang, H. K.; Sun, X. High-Dimensional Communication on Etchless Lithium Niobate Platform with Photonic Bound States in the Continuum. *Nat. Commun.* **2020**, *11* (1), 2602.

(38) Sun, W.; Liu, Y.; Qu, G.; Fan, Y.; Dai, W.; Wang, Y.; Song, Q.; Han, J.; Xiao, S. Lead Halide Perovskite Vortex Microlasers. *Nat. Commun.* **2020**, *11* (1), 4862.

(39) Wang, H.; He, Y.-M.; Chung, T.-H.; Hu, H.; Yu, Y.; Chen, S.; Ding, X.; Chen, M.-C.; Qin, J.; Yang, X. Towards Optimal Single-Photon Sources from Polarized Microcavities. *Nat. Photonics* **2019**, *13* (11), 770–775.

(40) Iff, O.; Buchinger, Q.; Moczała-Dusanowska, M.; Kamp, M.; Betzold, S.; Davanco, M.; Srinivasan, K.; Tongay, S.; Antón-Solanas, C.; Hofling, S. Purcell-Enhanced Single Photon Source Based on a Deterministically Placed WSe2 Monolayer Quantum Dot in a Circular Bragg Grating Cavity. *Nano Lett.* **2021**, *21* (11), 4715–4720.

(41) Yi, J.-M.; Smirnov, V.; Piao, X.; Hong, J.; Kollmann, H.; Silies, M.; Wang, W.; Groß, P.; Vogelgesang, R.; Park, N. Suppression of Radiative Damping and Enhancement of Second Harmonic Generation in Bull's Eye Nanoresonators. *ACS Nano* **2016**, *10* (1), 475–483.

(42) DeCrescent, R. A.; Wang, Z.; Imany, P.; Nam, S. W.; Mirin, R. P.; Silverman, K. L. Monolithic Polarizing Circular Dielectric Gratings on Bulk Substrates for Improved Photon Collection from InAs Quantum Dots. *Phys. Rev. Appl.* **2023**, *20* (6), 064013.

(43) Ge, Z.; Chung, T.; He, Y.-M.; Benyoucef, M.; Huo, Y. Polarized and Bright Telecom C-Band Single-Photon Source from InP-Based Quantum Dots Coupled to Elliptical Bragg Gratings. *Nano Lett.* **2024**, *24* (5), 1746–1752.

(44) Qu, L.; Bai, L.; Jin, C.; Liu, Q.; Wu, W.; Gao, B.; Li, J.; Cai, W.; Ren, M.; Xu, J. Giant Second Harmonic Generation from Membrane Metasurfaces. *Nano Lett.* **2022**, *22* (23), 9652–9657.

(45) Li, J.; Hu, G.; Shi, L.; He, N.; Li, D.; Shang, Q.; Zhang, Q.; Fu, H.; Zhou, L.; Xiong, W.; Guan, J.; Wang, J.; He, S.; Chen, L. Full-Color Enhanced Second Harmonic Generation Using Rainbow Trapping in Ultrathin Hyperbolic Metamaterials. *Nat. Commun.* **2021**, *12* (1), 6425.

(46) Fedotova, A.; Carletti, L.; Zilli, A.; Setzpfandt, F.; Staude, I.; Toma, A.; Finazzi, M.; De Angelis, C.; Pertsch, T.; Neshev, D. N. Lithium Niobate Meta-Optics. *ACS Photonics* **2022**, *9* (12), 3745–3763.

(47) Carletti, L.; Zilli, A.; Moia, F.; Toma, A.; Finazzi, M.; De Angelis, C.; Neshev, D. N.; Celebrano, M. Steering and Encoding the Polarization of the Second Harmonic in the Visible with a Monolithic LiNbO3 Metasurface. *ACS Photonics* **2021**, *8* (3), 731–737.

(48) Kim, K.-H.; Rim, W.-S. Anapole Resonances Facilitated by High-Index Contrast between Substrate and Dielectric Nanodisk Enhance Vacuum Ultraviolet Generation. *ACS Photonics* **2018**, *5* (12), 4769–4775.

(49) Huang, Z.; Lu, H.; Xiong, H.; Li, Y.; Chen, H.; Qiu, W.; Guan, H.; Dong, J.; Zhu, W.; Yu, J.; Luo, Y.; Zhang, J.; Chen, Z. Fano Resonance on Nanostructured Lithium Niobate for Highly Efficient and Tunable Second Harmonic Generation. *Nanomaterials* **2019**, *9* (1).

(50) Lin, Y.; Ye, Y.; Fang, Z.; Chen, B.; Zhang, H.; Yang, T.; Wei, Y.; Jin, Y.; Kong, F.; Peng, G. Efficient Second-Harmonic Generation of Quasi-Bound States in the Continuum in Lithium Niobate Thin

Film Enhanced by Bloch Surface Waves. *Nanophotonics* **2024**, *13* (13), 2335–2348.

**Acknowledgments**

**Funding:** This work was supported by National Natural Science Foundation of China (62022058, 12074252, 12192252); National Key R&D Program of China (2022YFA1205101, 2023YFA1407202); Shanghai Municipal Science and Technology Major Project (2019SHZDZX01-ZX06) and Yangyang Development Fund.

**Author contributions:** Y.Z. and X.C. conceived the idea and supervised the project. Z.L., J.F. and Z.H. developed the theory and performed the simulation. Z.L., X.Y. H.L. and S.L. designed and fabricated the samples. Z.L., X.Y. and Z.M. carried out the experiments. Z.L., Z.Y. and B.W. wrote the manuscript with the input from all authors.

**Competing interests:** The authors declare that they have no competing interests.

**Data and materials availability:** All data needed to evaluate the conclusions in the paper are present in the paper and/or the Supplementary Materials.